\documentstyle[prl,aps,psfig]{revtex}

\twocolumn

\newcommand{\beqa}{\begin{eqnarray}}
\newcommand{\eeqa}{\end{eqnarray}}
\newcommand{\non}{\nonumber}

\begin{document}   
\draft

\title{Perturbative description of nuclear double beta decay
transitions} 

\author{
D.R Bes$^{1,3}$,
O. Civitarese$^2$
and
N.N. Scoccola$^1$
}

\address{
$^1$Departamento de F\'{\i}sica, CNEA, Av.\ Libertador 8250,
(1429) Buenos Aires, Argentina.\\
$^2$Departamento de F\'\i sica, UNLP, C.C. 67, (1900)
La Plata, Argentina.\\
$^3$Instituto Universitario de C. Biom\'edicas, FURF, Sol\'\i s
453, (1078) Buenos Aires, Argentina.}
\maketitle

\begin{abstract}

A consistent treatment of intrinsic and collective
coordinates is applied to the calculation of matrix elements 
describing nuclear double beta decay transitions.
The method, which was developed for the case of nuclear
rotations, is adapted to include isospin and number of particles
degrees of freedom. It is shown
that the uncertainties found in most models, in dealing with these decay modes,
are largely due to the mixing of physical and spurious effects
in the treatment of isospin dependent interactions.

\end{abstract}


\renewcommand{\thefootnote}{\arabic{footnote}}

One can hardly overestimate the importance of the double beta
decay as a process explicitly linking the physics
of neutrinos with the nuclear structure \cite{ha:84,boh:92,mo:94}.   
Unfortunately,
the reliability of the theoretical predictions has been hampered
by unstabilities in the many-body BCS + RPA - type of treatments that
have been applied during the last 
decade \cite{vo:85,vo:86,ci:87,su:98}.     
An alternative approach
based on group theoretical methods has confirmed the existence
of a zero-energy state for certain values of the strength
of the proton-neutron, particle-particle, effective
interaction \cite{en:97,hi:97}. The appearance of such a state has been 
interpreted as a signature of a phase transition \cite{ci:97}.   

Here we take an alternative point of view
based on the fact  that 
the zero-energy state is a consequence of
the breakdown of the isospin symmetry implicit in the 
(separate) neutron and proton BCS solutions \cite{bm:75}. 

As similar to the 
case of deformed 
nuclei, the symmetry may be restored in the laboratory 
frame through the introduction
of collective coordinates. However,    
as different from previous cases dealing with collective coordinates, we 
must 
use here an interaction which does not conserve isospin,
in order to obtain
allowed matrix elements between $T\rightarrow (T-2)$ 
states:
{\em a central many-body problem that must be solved 
in double beta decay calculations is to disentangle 
unphysical isospin violations introduced by the theoretical treatment 
from physical effects   
due to the interaction}. 

In this letter we present the formalism
for the case of particles moving in a single j-shell
and coupled through a monopole pairing force. The hamiltonian
is not necessarily 
isoscalar. This simple 
case  involves all the complications 
associated with the collective treatment.
Moreover, the predictions may be compared with those of exact 
calculations \cite{hi:97} of
nuclear double beta decay transitions of the Fermi-type.
  
We define the operators
\beqa
S^+_v&=&\sum_{m>0}c^+_{vm}c^+_{v{\bar m}}\;;\;\;\;\;\;\;
S^+_{\perp}=\sum_{m>0}(c^+_{pm}c^+_{n{\bar m}}+c_{nm}^+
c^+_{p{\bar m}})\non ,\\
\tau_A&=&\frac{1}{2}(\tau_p+\tau_n)\;;\;\;\;\;
\tau_0=\frac{1}{2}(\tau_p-\tau_n)\;;\;\;\;\;
[\tau_{\bar 1},\tau_1]=\tau_0 \non , \\
T_A&=&{1\over2} \left(T_p+T_n\right)\;;\;\;\;\;\;
T_0={1\over2}\left(T_p-T_n\right)\;;\;\;\;\;\;[T_{\bar 1},
T_1]=-T_0\non ,\\
e_A&=&e_p+e_n\;;\;\;\;\;\;e_0=e_p-e_n  .
\eeqa
where  $\tau_v$ and 
$e_v$ 
are the number operators and 
single-particle energies, respectively, and $v=p,n$, ${\bar 1}=-1$.
The $T_v,T_{\pm 1}$ are the generators of collective 
rotations in gauge-and isospace.
 
The collective treatment appropriate for an isospin
conserving pairing interaction was introduced in refs.
\cite{bo:68}, \cite{gi:68}, \cite{du:71}.
The basic set of states associated with the collective sector
may be labeled by the four quantum numbers $|T_A;T,m,k>$,
where $T_A$ is the total number of pairs of particles.
We substitute
$M,T_0$ (the isospin projections  in the laboratory 
and intrinsic frames) by the quantum numbers 
$m\equiv \frac{1}{2}(T+M)$ and $k \equiv \frac{1}{2}(T+T_0)$,
respectively. 
We focus on states such that $m<<T$ and $k=0$. Hereof we drop the labels
$T_A,k$ from the collective states.
 
The introduction of collective degrees of freedom
is compensated through the appearance of the constraints
\beqa
\tau_z-T_z=0\;;\;\;\;\;\;\; (z=n,p,\pm 1) , \label{const}
\eeqa
 which express the fact that we can rotate the 
intrinsic system in one direction or the body in the opposite one without 
altering the physical situation \cite{bes:90a}. Physical states should be 
annihilated 
by the four constraints and physical operators should
commute with them.

Unphysical violations of the isospin symmetry  take 
place in the intrinsic frame, which
may be defined,
{\em for instance}, by the condition ${\bar S}_{\perp}=0$,
where the bar denotes the g.s. expectation value.
This condition  is precisely satisfied by performing a separate 
Bogoliubov transformation 
for protons and neutrons. 

Physical isotensor operators 
have to be transformed from the laboratory frame to the
intrinsic frame. In the case of the single-particle 
and pairing hamiltonians we
have
\beqa
H^{(lab)}_{sp}&=& e_A\tau_A+e_0(D^1_{00}\tau_0+D^1_{01}\tau_1
+D^1_{0{\bar 1}}\tau_{\bar 1})\non \\
&\approx& e_A\tau_A+e_0(\tau_0+d^+d+\frac{1}{T}
\tau_1\tau_{\bar 1})
\label{hspar}\\
H^{(lab)}_{p,0}&=&-g_0(S^+_pS_p+S^+_nS_n
+\frac{1}{2}S^+_{\perp}S_{\perp})\non \\
&\approx&-g_0(S^+_pS_p+S^+_nS_n)\non \\
&&+\frac{g_0}{T^2}[{\bar S}_n{\bar S}_p(\tau^2_1+
\tau^2_{\bar 1})+({\bar S}^2_n+{\bar S}^2_p)
\tau_1\tau_{\bar 1}]\label{hp0}\\
H^{(lab)}_{p,1}&=&-g_1[D^1_{00}(S^+_pS_p-S^+_nS_n)\non \\
\!\!\!\!\!\!\!\!&\!\!\!\!\!\!\!\!\!\!\!&\!\!\!\!\!\!
-\frac{D^1_{01}}{\sqrt{2}}(S^+_pS_{\perp}
+S^+_{\perp}S_n)\,+\,\frac{D^1_{0{\bar 1}}}{\sqrt{2}}(S^+_nS_{\perp}+
S^+_{\perp}S_p)]\non \\
&\approx&-g_1(S^+_pS_p-S^+_nS_n)\non \\
&&+\frac{g_1}{T}({\bar S}_p^2-{\bar S}_n^2)
(d^+d+\frac{1}{T}\tau_1\tau_{\bar 1}) \label{hp1}, \\
H^{(lab)}_{p,2}&=&-g_2\{D^2_{00}(S^+_pS_p+S^+_nS_n-S^+_{\perp}
S_{\perp})\non \\
\!\!\!\!\!\!\!\!\!\!\!\!&\!\!\!\!\!\!\!\!\!\!\!\!&\!\!\!\!\!\!\!\!\!\!\!\!
-\sqrt{\frac{3}{2}}[D^2_{01}(S^+_{\perp}S_n
-S^+_pS_{\perp})+D^2_{0{\bar 1}}(S^+_{\perp}S_p-S^+_n
S_{\perp})]\non  \\
&&+\sqrt{6}(D^2_{02}S^+_pS_n+D^2_{0{\bar 2}}
S^+_nS_p)\}\label{hp2} \\
\!\!\!\!\!\!\!\!&\!\!\!\!\!\!\!\!\!\approx&\!\!\!\!\!\!\!
-g_2[(S_p^+S_p+S^+_nS_n)+\frac{3}{T}{\bar S}_n{\bar S}_p
(\beta^{-4}d^2+\beta^4d^{+2})\non \\
\!\!\!\!\!\!\!\!\!\!\!\!&\!\!\!\!\!\!\!\!\!\!\!\!&\!\!\!\!\!\!\!\!\!\!\!\!
-({\bar S}^2_p+{\bar S}^2_n)
(\frac{3}{T}d^+d+\frac{1}{T^2}\tau_1
\tau_{\bar 1})-\frac{1}{T^2}{\bar S}_n{\bar S}_p(\tau_1^2+
\tau_{\bar 1}^2)] \non ,
\eeqa
plus null terms which are proportional to the constraints
(\ref{const}). We have kept  only the lowest
order terms in an expansion in powers of $T^{-1}$, assuming
$O({\bar S}_v)=T$ and $O(g_{\nu})=T^{-1}$. Use has been made
of the relation $S^+_{\perp}$ $ \approx$ $ -\frac{\sqrt{2}}{T}
({\bar S}_n\tau_1+{\bar S}_p\tau_{\bar 1})$. The $D^{\lambda}_{\mu \nu}$ 
have been expressed by means of
Marshalek's
generalization of the Holstein-Primakoff representation 
\cite{mar:75}, such that
the operators $\beta,d^+,d$ satisfy the relations $[d,d^+]=1$ 
and
\beqa
(\beta)^{2t}|T,m>& = &|T+t,m>\non , \\
(d^+)^t|T,m>& = &\sqrt{\frac{(m+t)!}{m!}}|T,m+t>.
\eeqa
It is easy to verify that the four components of the hamiltonian 
(\ref{hspar})-(\ref{hp2}) commute with the 
constraints (\ref{const}) and are therefore physical operators.

The contributions in  (\ref{hspar})-(\ref{hp2}) that are independent
of the  operators  $\beta,d^+,d,\tau_{\pm 1}$
add together to the two (proton-proton and neutron-neutron) 
pairing hamiltonians
in a single j-shell, namely 
\beqa
H_v=e_v\tau_v-g_vS_v^+S_v,
\eeqa
with interaction strengths $g_p=g_0+g_1+g_2$ and $g_n=g_0-g_1+g_2$.
These hamiltonians are separately treated within the BCS approximation.
Lagrange multiplier terms
$-\lambda_v(\tau_v-T_v)$ are added before solving the BCS equations.
This treatment yields the independent quasi-particle energy 
terms
$E_v {\nu}_v= {\frac {1}{2}} \Omega g_v {\nu}_v $,
where ${\nu}_v$ is the 
quasi-particle number operator for $v$-nucleons and $\Omega$ is half the
value of the shell degeneracy.
Within the RPA we may write
\beqa
E_p\nu_p+E_n\nu_n &\approx& -\frac{g_0+g_2}{T^2}\,
[{\bar S}_p{\bar S}_n(\tau^2_1+\tau^2_{\bar 1})\non \\
&&+(T^2+{\bar S}_n^2+{\bar S}_p^2)\tau_1\tau_{\bar 1}].
\label{hsp}  
\eeqa

Adding these contributions to the remaining 
terms in (\ref{hspar})-(\ref{hp2}) (i.e., to the ones 
depending on the operators
$\beta,d^+,d,\tau_{\pm 1}$ ), one obtains for the
proton-neutron sector of the spectrum (to leading order in $T^{-1}$)
\beqa
H^{(lab)}_{(sp+p)}&=& {\bar H}+\omega_dd^+d+H_2  \label{hfin} , \\
{\bar H}&=&e_v{\bar \tau}_v-g_v{\bar S}^2_v\non ,  \\
\omega_d&=& e_0+\frac{g_1}{T}({\bar S}_p^2-{\bar S}_n^2)
+\frac{3g_2}{T}({\bar S}^2_p+{\bar S}^2_n) \non, \\
\!\!\!\!\!\!\!\!\!\!\!\!\!\!\!\!\!\!&\!\!\!\!\!\! 
\!\!\!\!\!\!\!\!\!\!\!\!&\!\!\!\!\!\! \!\!\!\!\!\!\!\!\!\!
<T-2,m-2|H_2|T,m>=-\frac{3g_2}{T}{\bar S}_p{\bar S}_n\sqrt{m(m-1)}
\non ,
\eeqa
plus null terms. 

To leading order, the isospin operators in (\ref{hspar}) -
(\ref{hp2}) and in (\ref{hsp}) have
a boson structure since
 $[\tau_{\bar 1},(-\tau_1  )]\approx T$ 
and $\tau_{\bar 1}$ annihilates
the state with ${\bar \tau}_0=-T$. This is precisely the phonon that yields 
a zero frequency root for isoscalar hamiltonians
within a naive RPA \cite{hi:97}. In the present treatment this
phonon has disappeared 
from the final physical hamiltonian (10) to become part of the constraints 
(2). 
Nevertheless this degree of freedom 
must be taken into account in higher orders of the expansion in
powers of $T^{-1}$, for instance through the BRST procedure 
\cite{brst:74}, as     
applied  to many-body problems in \cite{bes:90a}, and
to the particular case of high angular momentum in 
\cite{bes:90b}.

The spectrum of the system is ordered into collective bands associated with 
particular values of the total number of particles and the isospin 
$(T \leq T_A)$. The energy of the band head is given by the BCS expectation
value $ {\bar H} $. 
The different members of each band are labeled by the quantum
number $m$ and are separated by the distance $\omega_d$, which includes 
the difference between the proton and the neutron single-particle energy
$e_0$. There is an interband interaction which mixes different values of $T$
but it conserves the projection $M$ in the laboratory frame
(cf. eq. (\ref{hfin})).

  From the point of view of the expansion in powers of $T^{-1}$, the interband
interaction is of the same order (O(1)) as the distance between the states
that are mixed by it. Nevertheless, in the following  we continue applying 
perturbation theory and we  also request that $|g_2|<g_v$.

In the calculations that we report in this paper we  
assume $g_p=g_n=g$.
The excitation energy $\omega_d$ is displayed for the cases $j=\frac{9}{2}$, 
$T_A=5$, $T=3$,  $e_0=0.8$~MeV, $g=0.4$~MeV  and 
$j=\frac{19}{2}$, $T_A=10$, $T=4$, $e_0=0.63$~MeV, $g=0.2$~MeV  as
 functions of the ratio $g_2/g$ (upper boxes of 
fig. 1). 
We predict the exact results for 
$g_2=0$ and very satisfactory ones for the 
other values, in spite of the 
fact that for these results we have neglected the interband interaction.

The strong current that appears in the weak hamiltonian is proportional to 
the isospin operator
\beqa
\beta_-&=&-\sqrt{2}\tau_1^{(lab)}=-\sqrt{2} (D^1_{11}\tau_1+D^1_{10}\tau_0
+D^1_{1{\bar 1}})\non \\
&\approx & \sqrt{2T}\,d^+\,+\,\mbox{null operator}.
\eeqa

The matrix element of double beta decay transitions, which for the
present case correspond to pure Fermi transitions 
(cf.  
\cite{hi:97}), 
is proportional to the product of the two matrix
elements
\beqa
M_1&=& <T,1|\beta_-|T,0>\approx \sqrt{2T}
\non\\
M_2&=&<(T-2),0|\beta_-|T,1>\non \\
&\approx &-\frac{2\sqrt{T}<(T-2),0|H_2|T,2>}
{{\bar H}(T,2)-{\bar H}((T-2),0)},
\eeqa   
These matrix elements are displayed in the lower boxes of fig. 1
 for the same 
parameters as in the upper boxes. The 
expression for the interband matrix element
in (\ref{hfin}) does not distinguish whether the r.h.s. 
should be calculated for the initial or the final value of
$T$, since it is valid for $T>>1$. Therefore,   
 the
effective interband matrix element has been chosen as the geometric 
average of the values obtained 
for 
each of the two connected bands. 

Fig. 2 displays Fermi double beta decay matrix elements, corresponding to 
transitions from the
initial to the final ground states. It has been calculated 
using the expression
\beqa
M_{2v}=\frac{M_1M_2}{\omega_d+\Delta}
\eeqa
where the energy released $\Delta$ has been taken to be
$0.5$ MeV, as in \cite{hi:97}. 
In addition to the exact and perturbative values of these
matrix elements, we have included in this figure the results
obtained by using some other approximations. The exact result
shows  the suppression of
the matrix element around the point where the strength of
the proton-neutron symmetry breaking interaction approaches
the value of the fully symmetric interaction. This
result is reproduced  both in the naive QRPA and in the 
perturbative approach.
The other approximation badly misses this cancellation.
A detailed comparison between the results of exact, naive QRPA and
renormalized QRPA (RQRPA) calculations can be found in  \cite{hi:97}.
It is worth to note that in the perturbative approach the
corresponding sum rule (Ikeda's sum rule) is exactly observed.
This is  not the case of other approaches, like in the
case of the RQRPA of \cite{toi:95}, where the
sum rule is violated. One can easily understand this failure
of the RQRPA approach, since it badly mixes-up terms in
defining the components of the equation of motion. Similar
conclusions about the validity of the RQRPA method are reported in
\cite{en:97}. The perturbative approach, as seen in figs. 1 and 2,
not only reproduces exact results very satisfactorily but it also
gives some insight about the mechanism responsible for the
suppression of the matrix elements. As found in the calculations,
the value of the matrix element $M_2$ depends critically on the
strength of symmetry breaking terms of the interaction between
protons and neutrons. The terms are proportional to $g_2$, as shown before.
On the other hand, the values of $M_1$ are not very much dependent
on this interaction. Finally, it should be observed that the
point where the excitation energy vanishes and the point where
the symmetry is completely restored are different (cf. fig. 1). This result,
also obtained in the exact diagonalization of the full
hamiltonian, cannot be reproduced 
by other means as shown in  \cite{hi:97}.  
Further details will be presented in a longer publication,
in which the extension of the formalism to include any number 
of non-degenerate 
j-shells has been performed and an isotensor isospin 
interaction has been included \cite{bes:98}.

In conclusion, it is found that a correct treatment of collective
effects induced by isospin dependent residual interactions in a
superfluid system is feasible:  physical effects due to
the isospin
symmetry-breaking terms in the hamiltonian are obtained even
in the presence of 
the BCS mean field built upon separate proton and neutron pairing
interactions. The definition
of intrinsic and collective coordinates and their separation 
guarantees that the isospin symmetry is restored and that
spurious contributions to the wave functions are decoupled from physical
ones. Particularly, the problem of the unstabilities found in 
the standard proton-neutron QRPA are avoided by the explicit
elimination of the zero frequency mode from the physical
spectrum but keeping it in the perturbative expansion.
The appearance of this mode cannot be
avoided by the inclusion of higher order terms in the
QRPA expansion or by any other ad-hoc renormalization procedure,
like the renormalized QRPA of \cite{toi:95}, once the BCS
procedure is adopted for the separate treatment of proton
and neutron pairing correlations.

 The results shown in this letter
are very encouraging, in spite of the fact that we have not used
very large values of $T$. We shall discuss the extension of the
formalism to cases such that $T<<O({\bar S}_v)$ and its application
to Gamow-Teller transitions in a forthcoming publication.

The authors are fellows of the CONICET, Argentina.
(O.C.) acknowledges the grant PICT0079 of the ANPCYT, 
Argentina; (D.R.B and N.N.S), grants from Fundaci\'on Antorchas.


\begin{figure}[tpb]
\centerline{\psfig{figure=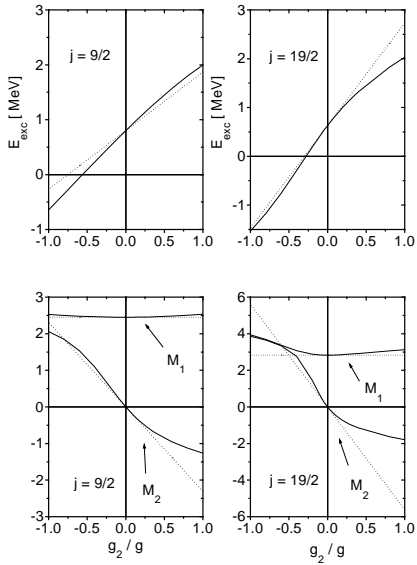,height=9cm}}
\protect\caption{\it 
Excitation Energy and Transition Matrix Elements. 
Exact (solid lines) and Perturbative (dotted lines) results
for the excitation energy (upper boxes) and transition
matrix elements $M_1$ and $M_2$ (lower boxes) corresponding
to the two different sets of parameters ($j=9/2$ and 
$j=19/2$) discussed in the text. }
\label{fig.1}
\end{figure}
\begin{figure}[tpb]
\centerline{\psfig{figure=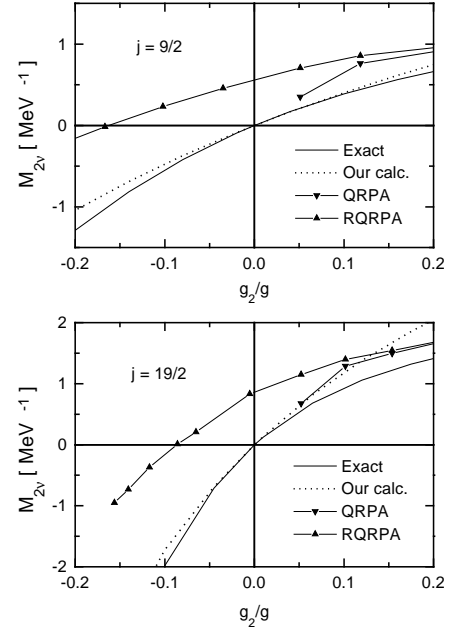,height=10cm}}
\protect\caption{\it
Matrix elements for Fermi double beta decay 
transitions calculated in several different approximations.
The meaning of approximations denoted as 
Quasiparticle Random Phase Approximation
(QRPA) and Renormalized Quasiparticle Random Phase
Approximation is explained in the text. Exact and
Perturbative results for the
two sets of parameters utilized in the calculations are 
indicated by solid and dotted lines, respectively. }
\label{fig.2}
\end{figure}

\end{document}